# Lepton polarization asymmetry in $B^*_{s,d} \to \mu^+\mu^-$ with new $Z'$ couplings

**P. Maji[*], S. Mahata, S. Biswas and S. Sahoo[†]**

Department of Physics, National Institute of Technology Durgapur

Durgapur-713209, West Bengal, India

[*]E-mail: majipriya@gmail.com, [†]E-mail: sukadevsahoo@yahoo.com

**Abstract**

We study the effect of non-universal $Z'$ boson on the rare leptonic decay modes of $B^*_s$ mesons, mediated by $b \to sll$ quark transition. Rare $B$ decays are sensitive to various new physics operators. As $B$ mesons are composite particles, such decay modes of the excited states are ideal for probing new physics beyond the standard model. We have constrained our model parameters from $B_{s,d} - \bar{B}_{s,d}$ mixing data to predict the lepton polarization asymmetry of $B^*_{s,d} \to \mu^+\mu^-$ channel. It is found that this asymmetry deviates from its SM prediction along with the $Z'$ effect.

**Keywords:** Rare B decays; $Z'$ boson; new physics; lepton polarization asymmetry

## I. Introduction

The rare decays of $B$ mesons are excellent testing ground to test the standard model (SM) precisely as well as to search for new physics (NP) lying beyond it. Recently, some unexpected phenomena in various $B$ meson decays are being observed in several experiments. Few of the observables are branching ratio of $B_s \to \varphi\mu^+\mu^-$ decay [1], angular observable $P'_5$ in $B \to K^*\mu^+\mu^-$ decay [2], lepton flavour non-universality parameters $R_{K^{(*)}}$ [3-6], $R_{D^{(*)}}$ [7, 8] etc. The fact that these observables show significant deviation around 3σ from their SM values which declare them as anomalies in recent time. To find the possible solutions, scientists extend their ideas beyond the SM which point towards the presence of NP. However, many well motivated extensions of the SM predict a branching ratio large enough to be seen at hadron colliders. Such NP models are like leptoquark [9-12], 2HDM [13-16], non-universal $Z'$ [17-19], fermion fourth generation [20-25] etc. These are being examined thoroughly to see whether they could explain the recent anomalies.

In this paper, we are interested to study the heavy-light systems like the $(b\bar{q})$ mesons which have a rich spectrum of excited states [26-28]. We are mainly concerned about the decay $B^*_{s,d} \to l^+l^-$ which includes $(b \to s,d)ll$ flavor-changing neutral-current (FCNC) transition. The excited mesons $B^*_{s,d}$ are unstable under electromagnetic and strong interactions and possess narrow width with corresponding lifetime of the order of $10^{-17}$s. The $B^*_{s,d} \to l^+l^-$ decays are sensitive to short-distance structure of $\Delta B = 1$ transitions. Some theoretical studies are being done in refs. [29-31] regarding these decay channels. The authors of ref. [30] have proposed a novel method to study FCNCs in the $B^*_{s,d} \to e^+e^-$ transition and predicted the branching ratio to be $BR(B_{s,d}^* \to e^+ e^-) = 0.98 \times 10^{-11}$. In ref. [31] $B_{s,d}^* \to l^+ l^-$ decay modes have been studied in the SM and the branching ratio has been predicted as $BR^{SM}(B_{s,d}^* \to l^+ l^-) = (0.7 - 2.2) \times 10^{-11}$ for decay width $\Gamma = 0.10(5)$ keV, irrespective of the lepton flavor. In our previous work [31], we have studied $B_{s,d}^* \to l^+ l^-$ ($l = \mu, e$) decay in $Z'$ model and predicted the branching ratio as $BR^{SM+Z'}(B_s^* \to l^+ l^-) = (1.5 - 2.2) \times 10^{-11}$ and $BR^{SM+Z'}(B_d^* \to l^+ l^-) = (1.7 - 2.2) \times 10^{-13}$.

Theoretical investigation of longitudinal lepton polarization asymmetry ($A_{P_L}$) is found to be more clean compared to the branching ratio of this decay channel as the observable $A_{P_L}$ is independent of the total width of $B^*$ meson which is not confirmed theoretically or experimentally. In this paper, we investigate the rare leptonic decay process $B_s^* \to l^+ l^-$ ($l = \mu$), mediated by $b \to s$ FCNC transitions. We first predict $A_{P_L}$ within the SM and then analyse its sensitivity to the non-universal $Z'$ model [32] which is an extension of SM with an extra $U(1)'$ symmetry. The main attraction of this NP model is that FCNC transitions could occur at tree level due to the off-diagonal couplings of non-universal $Z'$ with fermions, which is not allowed under SM consideration. The relation between the electroweak interaction eigenstates and mass eigenstates induces GIM mechanism within SM due to which flavor changing neutral interaction (FCNI) becomes forbidden at tree level. However, the relation between the electroweak interaction eigenstates of NP and the mass eigenstates is not same as of the SM. In such a situation, $Z'$ model could allow the tree level FCNC $b \to sll$ transitions. $B_s$ mixing and $B_s \to \mu^+ \mu^-$ data are highly favoured to determine the FCNC $b - s - Z'$ coupling. Another essential coupling is $\mu - \mu - Z'$ which is assumed same as the SM Z boson. As, $B_s^* \to l^+ l^-$ decay modes are not observed experimentally till now, so these decays are expected to be used to test the flavor sector of the SM and search for NP.

The paper is organized as follows. In section II, we discuss the effective Hamiltonian for $B_q^* \to l^+ l^-$ decay processes and give the formulation for lepton polarization asymmetry. The description of non-universal $Z'$ model is given in section III. In section IV, we discuss the contribution of $Z'$ boson in this channel. Section V contains the constraints coming from parameter space of $B_q - \bar{B}_q$ mixing data. Section VI is followed by concluding remarks.

## II. $B_q^* \to l^+ l^-$ decay processes in SM

The effective Hamiltonian for $B_q^* \to l^+ l^-$ decay induced by $b \to q l^+ l^-$ transition is given within the SM as [29, 33-36]

$$\mathcal{H}^{SM} = -\frac{G_F}{\sqrt{2}\pi} V_{tq}^* V_{tb} \left[ \sum_{i=1}^{6} C_i(\mu) \mathcal{O}_i(\mu) + C_7 \frac{e}{16\pi^2} [\bar{s}\sigma_{\mu\nu}(m_q P_L + m_b P_R)b] F^{\mu\nu} + C_9 \frac{\alpha}{4\pi} (\bar{q}\gamma^\mu P_L b)(\bar{\mu}\gamma_\mu \mu) + C_{10} \frac{\alpha}{4\pi} (\bar{q}\gamma^\mu P_L b)(\bar{\mu}\gamma_\mu \gamma_5 \mu) \right], \quad (1)$$

The vector meson $B^*$ has the same quark content as pseudo-scalar $B$ meson but with the state $J^P = 1^-$. Leptonic channels of $B$ meson suffer from large helicity suppression within the SM whereas $B_q^* \to l^+ l^-$ channels are free from such effects. This partially contributes for the shorter lifetime of $B^*$ meson. The SM amplitude for $B_q^* \to l^+ l^-$ decay channel is given as

$$\mathcal{M} = \frac{G_F}{2\sqrt{2}} V_{tb} V_{tq}^* \frac{\alpha}{\pi} \Big[\Big(m_{B_q^*} f_{B_q^*} C_9 + 2 f_{B_q^*}^T m_b C_7\Big) \bar{l} \not{\in} l + m_{B_q^*} f_{B_q^*} C_{10} \bar{l} \not{\in} \gamma_5 l -$$
$$8\pi^2 \frac{1}{q^2} \sum_{i=1}^{6,8} C_i \langle 0|T_i^\mu|B_q^*(p,\in)\rangle \bar{l}\gamma_\mu l\Big], \qquad (2)$$

where $G_F$ is the Fermi coupling constant, $\in$ is the polarization vector and $f_{B_q^*}$ is decay constant of $B_q^*$ meson. $C_i$'s are the Wilson coefficients of the weak Hamiltonian for $\Delta B = 1$ processes [37-39] evaluated at the $b$ quark mass scale [40] in the next-to-next-leading order. The coefficients $C_{9,10}$ are related to the short-distance semileptonic operators and $C_7$ is the coefficient of the electromagnetic penguin operator.

The matrix elements of the quark level operators are related to $B_q^*$ meson decay constants as follows:

$$\langle 0|\bar{q}\gamma^\mu b|B_q^*(p_{B_q^*},\in)\rangle = f_{B_q^*} m_{B_q^*} \in^\mu, \qquad (3)$$

$$\langle 0|\bar{q}\sigma^{\mu\vartheta} b|B_q^*(p_{B_q^*},\in)\rangle = -i f_{B_q^*}^T \Big(p_{B_q^*}^\mu \in^\vartheta - \in^\mu p_{B_q^*}^\vartheta\Big), \qquad (4)$$

$$\langle 0|\bar{q}\gamma^\mu \gamma_5 b|B_q^*(p_{B_q^*},\in)\rangle = 0, \qquad (5)$$

$$\langle 0|\bar{q}\gamma^\mu \gamma_5 b|B_q(p_B)\rangle = -i f_{B_q} p_{B_q}^\mu. \qquad (6)$$

The first two matrix elements comes from nonperturbative contributions where $f_{B_q^*}^T$ depends on the renormalization scale. In heavy quark limit, $f_{B_q^*}$ are related to $f_{B_q}$ as [40-42]

$$f_{B_q^*} = f_{B_q}\Big(1 - \frac{2\alpha}{3\pi}\Big), \qquad (7)$$

and $f_{B_q^*}^T$ are related to $f_{B_q}$ as

$$f_{B_q^*}^T = f_{B_q}\Big[1 + \frac{2\alpha}{3\pi}\Big(\log\Big(\frac{m_b}{\mu}\Big) - 1\Big)\Big]. \qquad (8)$$

The corresponding decay width for $B_s^* \to l^+ l^-$ decays is given as

$$\Gamma = \frac{G_F^2 \alpha^2 |V_{tb}V_{tq}^*|^2}{96\pi^3} m_{B_q^*}^3 f_{B_q^*}^2 \Bigg[\Bigg|C_9^{eff}\big(m_{B_q^*}^2\big) + 2\frac{m_b f_{B_q^*}^T}{m_{B_q^*} f_{B_q^*}} C_7^{eff}(m_{B_q^*}^2)\Bigg|^2 + |C_{10}|^2\Bigg]. \qquad (9)$$

The unit longitudinal polarization four-vector in the rest frame the lepton ($l^+$ or $l^-$) is defined as

$$\bar{s}_{l^\pm}^\alpha = \Big(0, \pm \frac{\vec{p}_l}{|\vec{p}_l|}\Big)$$

In the dilepton rest frame i.e. $B_s^*$ meson rest frame, these unit vectors will take the following form

$$s_{l^\pm}^\alpha = \left(\frac{|\vec{p}_l|}{m_l}, \pm\frac{E_l}{m_l}\frac{\vec{p}_l}{|\vec{p}_l|}\right),$$

where $E_l$, $\vec{p}_l$ and $m_l$ are the energy, momentum and mass of the lepton respectively.

Now, the two longitudinal polarization asymmetries $A_{P_L}^+$ for $l^+$ and $A_{P_L}^-$ for $l^-$ are given as

$$A_{P_L}^\pm = \frac{[\Gamma(s_{l^-}, s_{l^+}) + \Gamma(\mp s_{l^-}, \pm s_{l^+})] - [\Gamma(\pm s_{l^-}, \mp s_{l^+}) + \Gamma(-s_{l^-}, -s_{l^+})]}{[\Gamma(s_{l^-}, s_{l^+}) + \Gamma(\mp s_{l^-}, \pm s_{l^+})] + [\Gamma(\pm s_{l^-}, \mp s_{l^+}) + \Gamma(-s_{l^-}, -s_{l^+})]}, \quad (10)$$

where $s_{l^+}$ and $s_{l^-}$ are spin projections of the lepton. The decay rates for same spin and opposite spin leptons could be found in Ref. [43]

The final expression for polarization asymmetry can be written as

$$A_{P_L}^\pm|_{SM} = \mp \frac{2\sqrt{1 - \frac{4m_l^2}{m_{B_q^*}^2}}\, \mathrm{Re}\left[\left(C_9 + \frac{2m_b f_{B_q^*}^T}{m_{B_q^*} f_{B_q^*}} C_7\right) C_{10}^*\right]}{\left(1 + \frac{2m_l^2}{m_{B_q^*}^2}\right)\left|C_9 + \frac{2m_b f_{B_q^*}^T}{m_{B_q^*} f_{B_q^*}} C_7\right|^2 + \left(1 - \frac{4m_l^2}{m_{B_q^*}^2}\right)|C_{10}|^2} \quad (11)$$

### III.  Formalism of the $Z'$ model

In the $Z'$ model considered here, the FCNC $b - s - Z'$ coupling is related to the flavor diagonal couplings $qqZ'$. We follow the formalism given in Ref. [44] to construct the model. For simplicity we neglect the mixing angle between $Z$ and $Z'$ which is $\mathcal{O}(10^{-4})$. The current corresponding to the additional $U(1)$ gauge symmetry is

$$J_\mu^{(2)} = \sum_{i,j} \bar{\psi}_i \gamma_\mu \left[\epsilon_{\psi_{L_{ij}}}^{(2)} P_L + \epsilon_{\psi_{R_{ij}}}^{(2)} P_R\right]\psi_j, \quad (12)$$

where $P_{L,R}$ are projection operators and $\epsilon_{\psi_{L,R_{ij}}}^{(2)}$ are the chiral couplings of $Z'$ with fermions. We adopt the following considerations for our whole work

i) Up-type quarks are assumed flavor diagonal and family universal: $\epsilon_{L,R}^u = Q_{L,R}^u \mathbf{1}$, $\epsilon_{L,R}^e = Q_{L,R}^e \mathbf{1}$ and $\epsilon_L^\nu = Q_L^\nu \mathbf{1}$, where $\mathbf{1}$ is the $3 \times 3$ identity matrix and $Q_{L,R}^u, Q_{L,R}^e, Q_L^\nu$ are the chiral charges.

ii) Down-type quarks are interacting with $Z'$ as

$$\mathcal{L}_{NC}^{(2)} = -g_2 Z'_\mu (\bar{d}, \bar{s}, \bar{b})_I \gamma^\mu (\epsilon_L^d P_L + \epsilon_R^d P_R)\begin{pmatrix} d \\ s \\ b \end{pmatrix}_I,$$

where the subscript $I$ stands for interaction basis and $g_2$ is the $U(1)'$ coupling.

iii) To specify the RH chiral couplings of the down sector and the chiral couplings of the up sector, we assume

$$|Q_R^d| = |Q_L^d|, \qquad |Q_{L,R}^u| = |Q_{L,R}^d|.$$

iv) The chiral couplings of $Z'$ to leptons ($B_{ll}^{L,R}$) are assumed to have the form as the coupling of the SM $Z$ boson to leptons [17]

$$B_{ll}^L = T_{3l}^L - \sin^2\theta_W Q_l, \qquad B_{ll}^R = T_{3l}^R - \sin^2\theta_W Q_l$$

where $T_{3l}^L$ ($T_{3l}^R$) is the third component of weak isospin for the left (right) chiral components, $Q_l$ is the charge of leptons and $\theta_W$ is the weak mixing angle. For all the charged lepton families $T_{3l}^L = 1/2$ and $T_{3l}^R = 0$.

## IV. $B_{s,d}^* \to l^+ l^-$ decay processes in $Z'$ model

In the $Z'$ model, tree level $b \to q l^+ l^-$ transitions are allowed due to the presence of extra gauge boson. The effective Hamiltonian for such processes are given as [45, 46]

$$\mathcal{H}_{eff}^{Z'} = \frac{2G_F}{\sqrt{2}} V_{tb} V_{tq}^* \left[(\bar{s}b)_{V-A}\left(C_{ll}^{bq}(l\bar{l})_{V-A} + D_{ll}^{bq}(l\bar{l})_{V+A}\right)\right] \tag{13}$$

where $C_{ll}^{bq} = \left(\frac{g_2 M_Z}{g_1 M_{Z'}}\right)^2 \frac{B_{qb}^L B_{ll}^L}{V_{tb} V_{tq}^*}$; $D_{ll}^{bq} = \left(\frac{g_2 M_Z}{g_1 M_{Z'}}\right)^2 \frac{B_{qb}^L B_{ll}^R}{V_{tb} V_{tq}^*}$

and we parameterize the couplings as following

$$\rho_{qb}^L = \frac{g_2 M_Z}{g_1 M_{Z'}} B_{qb}^L = \frac{g_2 M_Z}{g_1 M_{Z'}} |B_{qb}^L| e^{i\phi_{qb}^L} \tag{14}$$

$$\rho_{ll}^{L,R} = \frac{g_2 M_Z}{g_1 M_{Z'}} B_{ll}^{L,R} \tag{15}$$

Comparing this Hamiltonian with the following equation

$$\mathcal{H}_{eff}^{Z'} = -\frac{G_F}{\sqrt{2}\pi} V_{tb} V_{tq}^* \left(\Lambda_{qb} C_9^{Z'} Q_9 + \Lambda_{qb} C_{10}^{Z'} Q_{10}\right) \tag{16}$$

we get

$$\Lambda_{qb} = \left(\frac{g_2 M_Z}{g_1 M_{Z'}}\right)^2 \frac{2\pi e^{i\phi_{qb}^L}}{\alpha V_{tb} V_{tq}^*}; \quad C_9^{Z'} = -|B_{qb}^L| S_{LL}; \quad C_{10}^{Z'} = |B_{qb}^L| T_{LL}. \tag{17}$$

$$S_{LL} = B_{ll}^L + B_{ll}^R \text{ and } T_{LL} = B_{ll}^L - B_{ll}^R$$

Now, we can have the effect of new $Z'$ boson to the decay parameter i.e. lepton polarization asymmetry by modifying the Wilson coefficients $C_9$ and $C_{10}$. The new form of $A_{P_L}$ will be

$$A^{\pm}_{P_L}|_{SM} = \mp \frac{2\sqrt{1-\frac{4m_l^2}{m_{B_q^*}^2}} Re\left[\left(C_9^{tot} + \frac{2m_b f_{B_q^*}^T}{m_{B_q^*} f_{B_q^*}} C_7\right) C_{10}^{tot*}\right]}{\left(1+\frac{2m_l^2}{m_{B_q^*}^2}\right)\left|C_9^{tot} + \frac{2m_b f_{B_q^*}^T}{m_{B_q^*} f_{B_q^*}} C_7\right|^2 + \left(1-\frac{4m_l^2}{m_{B_q^*}^2}\right)|C_{10}^{tot}|^2} \quad (18)$$

Here $C_9^{tot} = C_9 + \Lambda_{sb} C_9^{Z'}$; $C_{10}^{tot} = C_{10} + \Lambda_{sb} C_{10}^{Z'}$.

## V. Analysis

The mass difference of the $B_q$ system within the SM is given by [37]

$$\triangle m_{B_q}^{SM} = \frac{G_F^2}{6\pi^2} M_W^2 m_{B_q} f_{B_q}^2 (V_{tb} V_{tq}^*)^2 \eta_{2B} S_0(x_t) [\alpha_s(m_b)]^{-\frac{6}{23}} \left[1 + \frac{\alpha_s(m_b)}{4\pi} J_5\right] \mathfrak{B}_{B_q}(m_b) \quad (19)$$

where the 'Inami-Lim' loop function $S_0(x_t) = 2.463$, $x_t = (m_t/M_W)^2$, QCD correction parameter at NLO $\eta_{2B} = 0.551$, $J_5 = 1.627$ [37].

From the vacuum insertion method we get

$$\langle \bar{B}_q | O^{LL} | B_q \rangle = 8/3 \mathfrak{B}_{B_q} f_{B_q}^2 m_{B_q}^2$$

Here $\mathfrak{B}_{B_q}$ is the bag parameter, $\widehat{\mathfrak{B}}_{B_s} = 1.5159 \mathfrak{B}_{B_s} = 1.320 \pm 0.017 \pm 0.03$ and $\frac{\widehat{\mathfrak{B}}_{B_s}}{\widehat{\mathfrak{B}}_{B_d}} = 1.023 \pm 0.013 \pm 0.014$ [47]. $f_{B_s} = 230 \pm 30$ MeV is the decay constant of $B_s$ meson [36]. From the recent data of $\xi = 1.206 \pm 0.017 \equiv \frac{f_{B_s}\sqrt{\widehat{\mathfrak{B}}_{B_s}}}{f_{B_d}\sqrt{\widehat{\mathfrak{B}}_{B_d}}}$ [48], the decay constants can be determined with much smaller theoretical error. The error on $\triangle m$ within the SM can be evaluated from

$$\triangle m_{B_s}^{SM} = \triangle m_{B_d}^{SM} \xi \frac{m_{B_s}}{m_{B_d}} \frac{(1-\lambda^2)^2}{\lambda^2[(1-\bar{\rho})^2+\bar{\eta}^2]} \quad (20)$$

The Wolfenstein and other CKM parameters can be found in PDG 2020 [48].

After taking all the asymmetries in quadrature we obtain the SM prediction of mass difference as follows

$$\triangle m_{B_s}^{SM} = 18.952 \pm 5.497 \text{ ps}^{-1} \text{ and } \triangle m_{B_d}^{SM} = 0.672 \pm 0.120 \text{ ps}^{-1} \quad (21)$$

Whereas the experimental values are found as [48]

$$\triangle m_{B_s}^{exp} = 17.749 \pm 0.020 \text{ ps}^{-1} \text{ and } \triangle m_{B_d}^{exp} = 0.507 \pm 0.002 \text{ ps}^{-1} \quad (22)$$

So the effect of LH FCNC induced by $Z'$ is given by

$$\frac{\Delta m_{B_s}^{exp}}{\Delta m_{B_s}^{SM}} = \left|1 + 11.626 \times 10^5 (\rho_{sb}^L)^2 e^{2i\phi_{sb}^L}\right| = 0.937 \pm 0.271 \qquad (23)$$

$$\frac{\Delta m_{B_d}^{exp}}{\Delta m_{B_d}^{SM}} = \left|1 + 271 \times 10^5 (\rho_{db}^L)^2 e^{2i\phi_{db}^L}\right| = 0.753 \pm 0.137 \qquad (24)$$

where $\phi_{qb}^L$ is the new weak phase associated with the LH quark coupling $B_{qb}^L$.

We have shown the allowed parameters space of $(\rho_{sb}^L, \phi_{sb}^L)$ and $(\rho_{db}^L, \phi_{db}^L)$ varying the ratio of mass difference $\frac{\Delta m_{B_q}^{exp}}{\Delta m_{B_q}^{SM}}$ within $1\sigma$ allowed range in Figs. 1 and 2.

Here, Fig. 1 represents the constraint due to $B_s - \bar{B}_s$ missing and Fig. 2 for the $B_d - \bar{B}_d$ mixing. From these graphical representations we have obtained the constraints on $\rho_{sb}^L$ in $B_s - \bar{B}_s$ mixing as follows

$$0 \leq \rho_{sb}^L \leq 0.435 \times 10^{-3} \text{ for } 0° \leq \phi_{sb}^L \leq 180° \qquad (25)$$

Similarly, for $B_d - \bar{B}_d$ mixing case the bounds are

$$0.6 \times 10^{-4} \leq \rho_{db}^L \leq 1.2 \times 10^{-4} \text{ for } 60° \leq \phi_{db}^L \leq 120° \qquad (26)$$

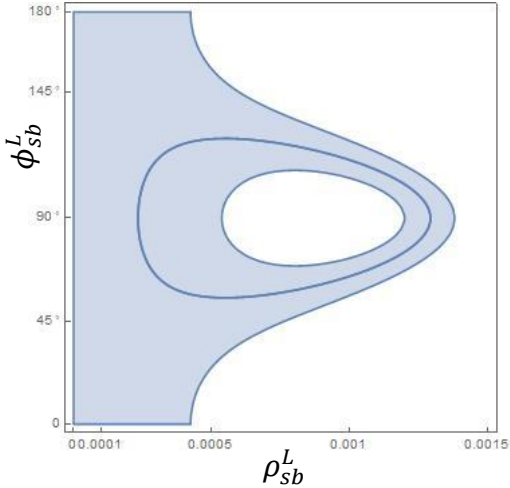
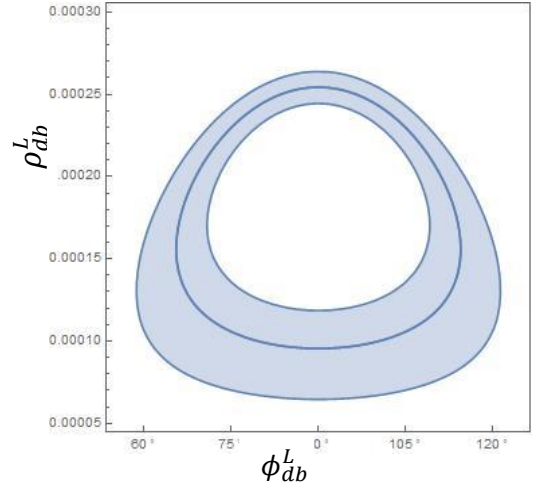

**Fig. 1:** Constraints on $\rho_{sb}^L$ and $\phi_{sb}^L$ from $B_s - \bar{B}_s$ mixing within $1\sigma$

**Fig. 2:** Constraints on $\rho_{db}^L$ and $\phi_{db}^L$ from $B_d - \bar{B}_d$ mixing within $1\sigma$

With all these constrains now we calculate the lepton polarization asymmetry for $B_{s,d}^* \to \mu^+\mu^-$ decay channels. Using the above values we get

**Table 1:** Lepton polarization asymmetry of $B_{s,d}^* \to \mu^+\mu^-$ channels

| Decay channels | $A_{P_L}^+|_{SM} = -A_{P_L}^-|_{SM}$ | $A_{P_L}^+|_{Z'} = -A_{P_L}^-|_{Z'}$ |
|---|---|---|
| $B_s^* \to \mu^+\mu^-$ | $-0.996 \pm 0.002$ | $-0.999 \pm 0.003$ |
| $B_d^* \to \mu^+\mu^-$ | $-0.995 \pm 0.001$ | $-0.990 \pm 0.001$ |

To calculate lepton polarization asymmetry in $Z'$ model, we consider the maximum values of coupling and weak phase from the bounds i.e. $\rho_{sb}^L = 0.435 \times 10^{-3}$ for $\phi_{sb}^L = 180°$ and $\rho_{db}^L = 1.2 \times 10^{-4}$ for $\phi_{db}^L = 120°$. It is seen from Table 1 that for $B_s^* \to \mu^+\mu^-$ channel, lepton polarization asymmetry increases from the SM and reaches to $\sim -1$ and for $B_d^* \to \mu^+\mu^-$ channel it is decreased from the SM prediction. If the lower ranges are taken for the calculation, results are seemed to be same. We have also plotted this observable varying the full range of model parameters and shown them in Figs. 3 and 4.

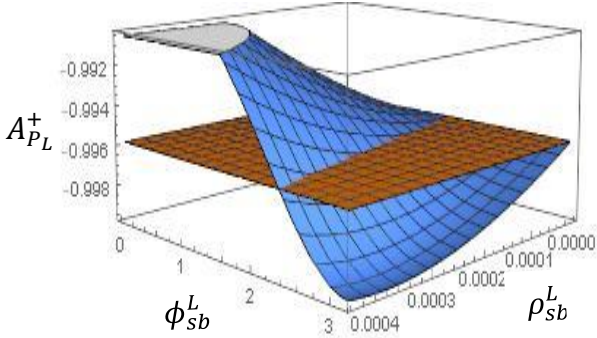 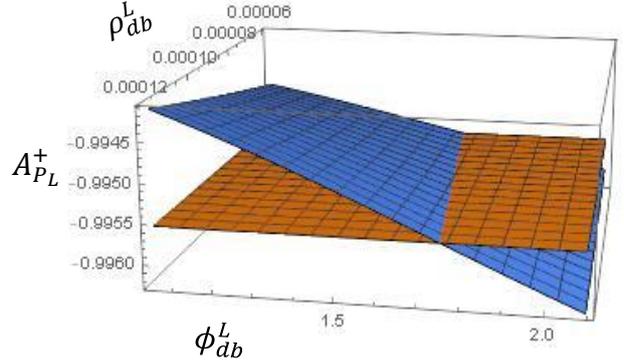

**Fig. 3:** Lepton polarization symmetry for $B_s^* \to \mu^+\mu^-$ channel.
Orange plate: SM
Blue plate: $Z'$ model

**Fig. 4:** Lepton polarization symmetry for $B_d^* \to \mu^+\mu^-$ channel.
Orange plate: SM
Blue plate: $Z'$ model

## VI.     Conclusion

In this paper, we have studied pure leptonic decay processes $B_{s,d}^* \to \mu^+\mu^-$ in the light of family non-universal $Z'$ model and estimated the muon longitudinal polarization asymmetry. These decays are highly dominated by radiative background and thus are not well measured. They do not suffer much from helicity suppression and sensitive to semileptonic operators. This fact could lead us to some extent of new physics at TeV scale. The lepton polarization asymmetry for the vector meson $B_s^*$ in SM is recently predicted in Ref. [43]. This observable is theoretically clean because it has only a very mild dependence on the vector meson decay constants. Here, we see that the asymmetry is deviated from their SM predictions. From the graphical representations, we notice that the asymmetry touches the SM for certain choice of model parameters. There are also several observations in the $b \to sll$ induced decay processes which do not agree with their SM prediction. All of these hint towards new physics beyond the SM. Recent values of $B - \bar{B}$ mixing are used to constraint $b - s - Z'$ couplings. As $Z'$ boson is not confirmed yet, further measurements are needed to confirm the couplings more precisely. Here, we have investigated the decay channels which are rare to achieve. We hope this study could be one step towards the search of new physics in future.

**Acknowledgement**

Maji is thankful to DST, Govt. of India for providing INSPIRE Fellowship (IF160115). Mahata and Biswas thank NIT Durgapur for providing fellowship. Sahoo is grateful to SERB, DST, Govt. of India for financial support through project (EMR/2015/000817).